# A possible role of Alpha Crucis in the astronomical landscape of Silbury Hill

**Amelia Carolina Sparavigna**
Politecnico di Torino

The paper is discussing a possible link between the construction of Silbury Hill, the prehistoric artificial mound near Avebury, and the observation of Alpha Crucis, the main star of the Crux constellation, which was slowly disappearing from the local sky due to the precession of the Earth's axis. For the discussion, we use simulations of the local astronomical landscape made by means of Stellarium software.

Silbury Hill is a prehistoric artificial mound near Avebury. Being 40 meters high, [1] it is the tallest prehistoric man-made mound in Europe [2] (Figure 1). The artificial hill was constructed in several stages, probably completed in around 2400 BC [3,4]. Since technical skills and a long-term control over works and resources were necessary for its building, the existence of a theocratic elite in the southern Britain was proposed [5], able of supervising the construction of the Hill. As told in [1], in an archaeological research that lasted from 1968 to 1970, Richard Atkinson, that undertook excavation at the Hill finding material suggesting a Neolithic date of the mound, indicated a date for its commencement close to 2750 BC [6].

What was the purpose of this hill? We can find some assertions in [1]. John Barret, in [7], notes that any ritual at Silbury Hill would have involved physically raising a few individuals far above the level of everyone else. These few individuals were therefore in a privileged position, possibly indicating an elite group, perhaps a priesthood, having some authority. Writer and prehistorian Michael Dames has proposed a theory of seasonal rituals, in an attempt to explain the purpose of the Hill and of its associated sites [8]. Paul Devereux observed that Silbury and its surrounding monuments appear to have been designed with a system of inter-related sightlines [1]. Jim Leary and David Field, after an overview of the site, concluded that the actual purpose of this artificial earth mound cannot be known and that the multiple and overlapping construction phases suggest that the process of building was probably the most important thing of all [9].

Let us consider the proposal of John C. Barret, that is viewing the hill as a place where a few individuals could have been in a privileged position and link this fact to the astronomical landscape of the site at the time of its construction. Let us also consider that the rising and setting of the stars is influenced by the precession of the Earth's axis (precession of equinoxes). Then, due to this phenomenon, ancient people could see some stars becoming circumpolar stars, that never set, and see other stars disappearing below the horizon forever. It was then important to have a platform, on which people had the possibility of freely observing the horizon, not impeded by nearby obstacles, such as the canopy of the trees.

As observed by Giulio Magli in [10], astronomy was a familiar presence in ancient sites. In the Mediterranean regions, besides those oriented to the sun and the moon, many sanctuaries existed probably "oriented to the brilliant stars of the southern sky" [10], such as those of the Crux-Centaurus group. These stars were among the stars which were "slowly disappearing from the Mediterranean sky due to precession [10,11]". At the same time, they were losing their importance for people living there. That the Crux-Centaurus group was important in the past for people of the Northern hemisphere is demonstrated by its relevant role in the ancient astrology and mythology. Ptolemy recognized the stars of the Crux as part of the constellation Centaurus, where "they marked the bottom of the hind legs and were thus, as were all the stars of the 'equine part', denoted as akin to Venus and Jupiter in influence" [12]. As explained in [12], "ancient identification with the bestial part of Centaurus, gives the bright stars of Crux similar associations to those of Bungula, on the hoof of the Centaur's right foreleg". This star was linked to the mythological figure of Cheiron by Eratosthenes as early as the 2nd century BC, and has been used to signify sacrifice, healing and the need to be healed [12].

We can imagine that these stars were important for people at northern latitudes too, such as those living near Avebury and Stonehenge. In fact, about 3000 BC, at Avebury, the stars of the Crux, in particular the Alpha Crucis (Acrux, the brightest star in the constellation, the 13th brightest star in the night sky), were important because indicating the South direction, where the South Pole of the Axis Mundi is under the horizon. As well explained in [13], this pole was sinking beneath the horizon due to the precession of the Earth's axis. In a very interesting discussion and interpretation, Nicholas Mann is linking Acrux and the Silbury Hill, telling that may be, the Hill was imagined as a sanctuary of these sinking stars: "Silbury Hill with its remarkable springs appears as the primal mound, the multi-tiered World Mountain, the ultimate symbol appropriate for a world centre and axis mundi between the worlds. Silbury Hill crowned the horizon in the place where the stars of the Southern Pole had entered the Underworld" [13].

Today, we are used to see the North Pole, almost coincident with Polaris, as a remarkable point of the night sky. However, 5,000 years ago, this was not so. The North Pole was marked by the faint Thuban in the constellation Draco, whereas, in the Avebury nighty sky, there were the bright stars of the Crux shining southward. It is then evident the importance of then and of the South direction for people who lived in this part England. In the Figure 2, we can see the winter sky in this direction simulated by Stellarium software (the use of which in archaeoastronomy was proposed in [14,15]), for the Silbury Hill latitude in 3,000 BC. If we consider the effect of the atmospheric refraction, Acrux appeared sliding on the horizon at least for an hour.

We can also use Stellarium to see the effect of precession simulating the sky at the latitude of the Silbury Hill, at the time of its construction. In the Figure 3, we show the snapshots of the simulation for 3000 BC, 2750 BC and 2500 BC. We have from the Figure 2 and the upper panel of Figure 3, that, about 3000 BC Acrux was above the astronomical horizon. In the middle panel of Figure 3, we can see that this star was, about 2750 BC, slightly below the astronomical horizon. In the lower panel of the same Figure, we can see it is below the horizon, no more visible for sure. As the simulation is showing, before and during the building of Silbury Hill, Alpha Crucis was disappearing below the horizon. May be, the Hill was built as an astronomical observatory, to observe this phenomenon.

Let us stress an important fact. In the panel of the Figure 3 showing the southern sky about 2750 BC, that is at the time when the building of Silbury Hill started, Acrux is slightly below the astronomical horizon, so we could image it were not visible. However, we have to consider an important fact, a fact that we have already mentioned for the Figure 2, the existence of the atmospheric refraction. This refraction is the deviation of light from a straight line as it passes through the atmosphere due to the variation in air density as a function of altitude. Astronomical refraction causes stars to appear higher in the sky than they are in reality. The refraction of the light from a star is zero in the zenith, less than one arc-minute at 45 degrees apparent altitude. However, it increases as altitude decreases, becoming of 35.4 arc-minutes at the horizon [16]. This corresponds to more than a half of degree. Then, a star slightly below the horizon is still visible due to the atmospheric refraction.

However, besides refraction, atmosphere has another effect on the light of stars. The photons coming from them are subjected to scattering and absorption from the molecules of atmosphere and then only bright stars are visible when their altitude is very low [13,17,18]. Due to this effect, let us use Stellarium with the simulation of the presence of atmosphere. In the Figure 4, we have two panels showing the Crux, simulated with and without atmosphere, around 3000 BC. In the following Figure 5, we can see that, around 2800 BC, Acrux was very faint, probably visible for very good and trained eyes of local stargazers. Around 2750 BC, in the simulation (Figure 6), Acrux is a dot near the line of the horizon. In this manner, in a relatively short period of time, the stargazers at the latitude of Silbury Hill had observed the Acrux as "entered the Underworld".

As a conclusion, we can imagine that people could have started constructing the hill as an astronomical platform, to have a free horizon for the observation of the stars of the Crux and of Alpha Crucis in particular. As time passed, the Crux was lower and closer to the astronomical horizon, and one of them even below it, but still visible due to atmospheric refraction. May be, people imagined that a higher platform could help them to have a better observation of these stars (this could explain the several stages of the building of Silbury Hill).

In fact, a platform (height h) allowed the builders of Silbury Hill to see a larger horizon. They could gain a visual angle α of about 0.10 degree for h = 40 m. We can find the angle using formulas as those given in https://en.wikipedia.org/wiki/Horizon. Approximately: sin α = h / sqrt(2hR), where R is the radius of the Earth. It is a small angle, when compared with the effect of precession, which is, on a hundred of years, of about 0.4 degrees (see the Figure 5 for instance). It is therefore possible that the stargazers at Silbury Hill had the possibility to compensate the effect of precession for a short period of time. However, they probably realized that it was a vain effort to continue in increasing the height of the Hill. In any case, the overlapping construction phases seem suggesting a link of the process of building to the evolution of the sky, in the effort to continue to see the sky as it was according to the lore of ancient times.

**Note**

In the previous layouts of the paper I have underestimated the gain of visual angle obtained by the height of the Hill. I revised accordingly the paper.

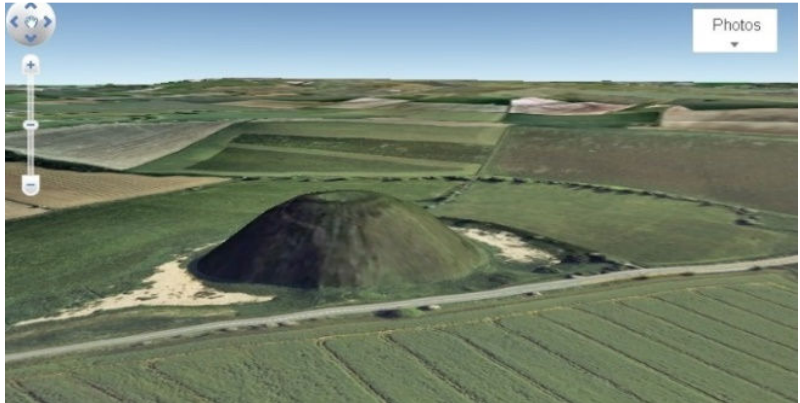
Figure 1: Silbury Hill in a Google Earth view.

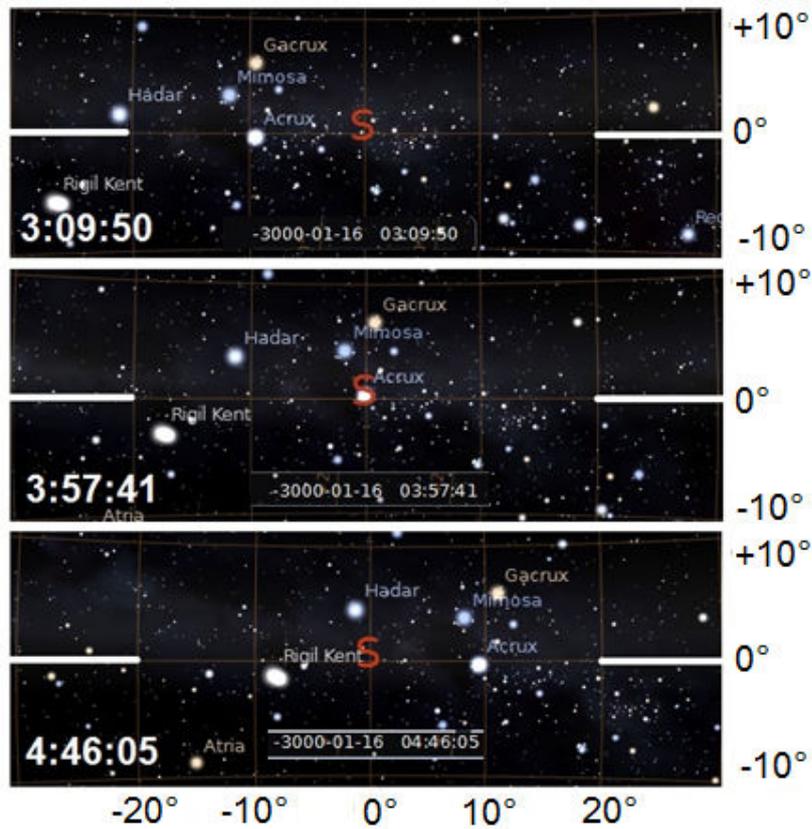
Figure 2: Simulation of a winter night of 3000 BC by means of Stellarium. The white line is the astronomical horizon. Angles are given in an azimuthal frame of reference. The stars of the Crux, in particular Alpha Crucis (Acrux) are sliding on the horizon. If we add the effect of atmospheric refraction, `which` causes stars to appear higher in the sky than they are in reality, Acrux could have been observed for at least an hour.

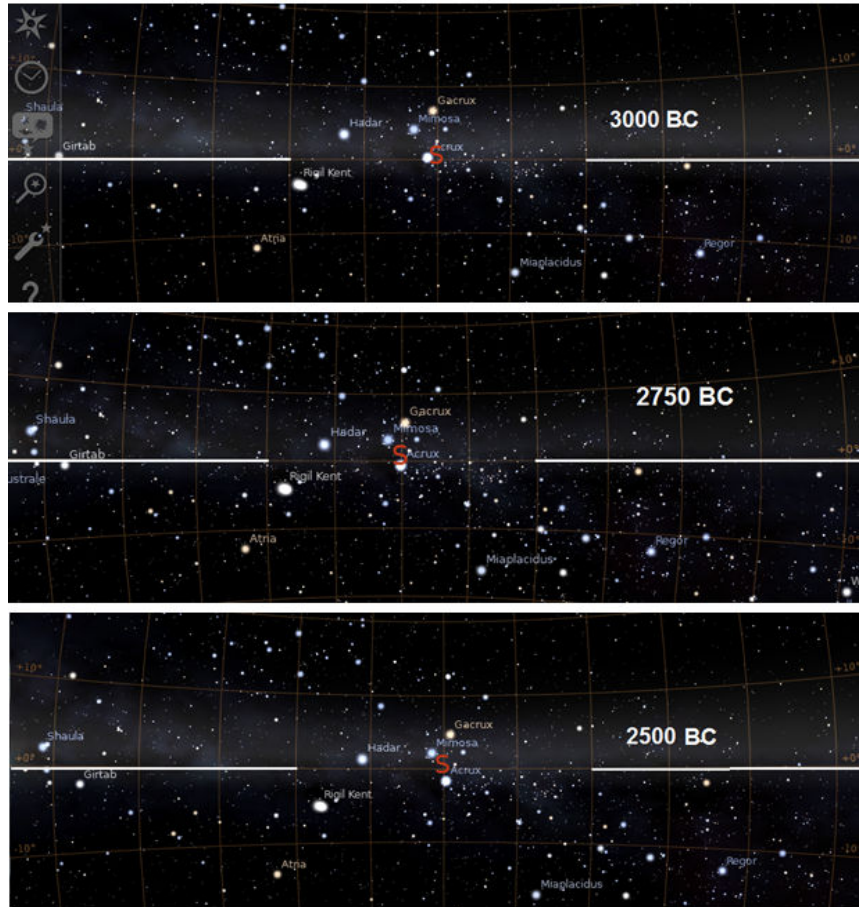

Figure 3: The stars of Crux-Centaurus (Acrux is Alpha Crucis) in 3000, 2750 and 2500 BC as given by Stellarium (atmosphere off) for the Silbury Hill location. The white line is the astronomical horizon. In 2750 BC, Acrux is just below the horizon. Probably, it was very faint but still visible due to the atmospheric refraction.

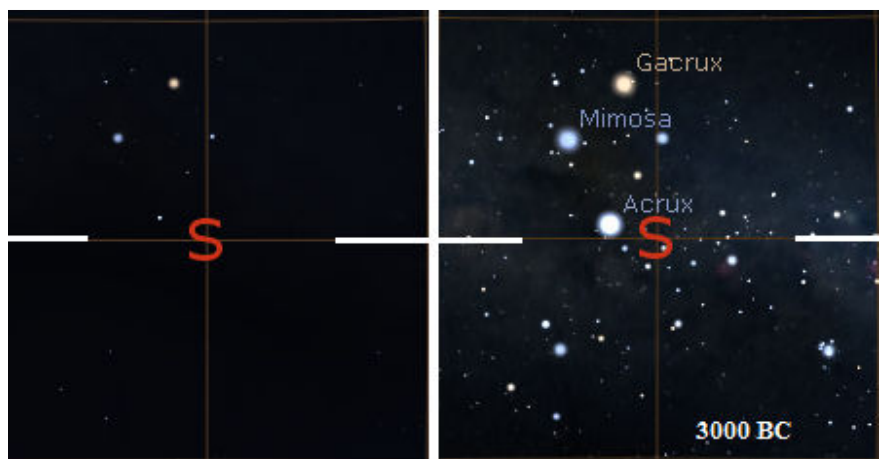

Figure 4: Stellarium simulates the effect of atmosphere, as we can see in the left panel, which is showing the stars of the Crux, around 3000 BC.

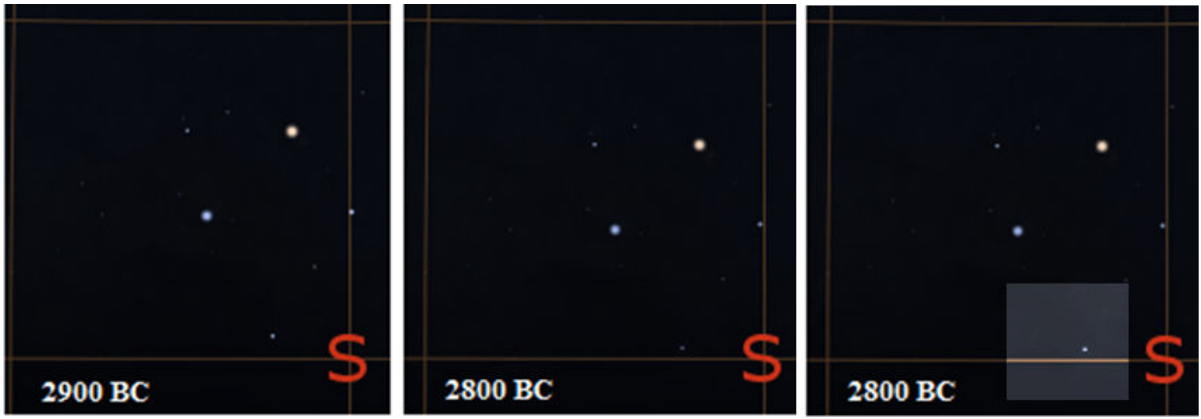

Figure 5: The left and middle panels are showing the results given by Stellarium (atmosphere on). Around 2800 BC, Acrux is very faint, but it is visible in the simulation (in the right panel I have adjusted the brightness/contrast of the area where there is the star).

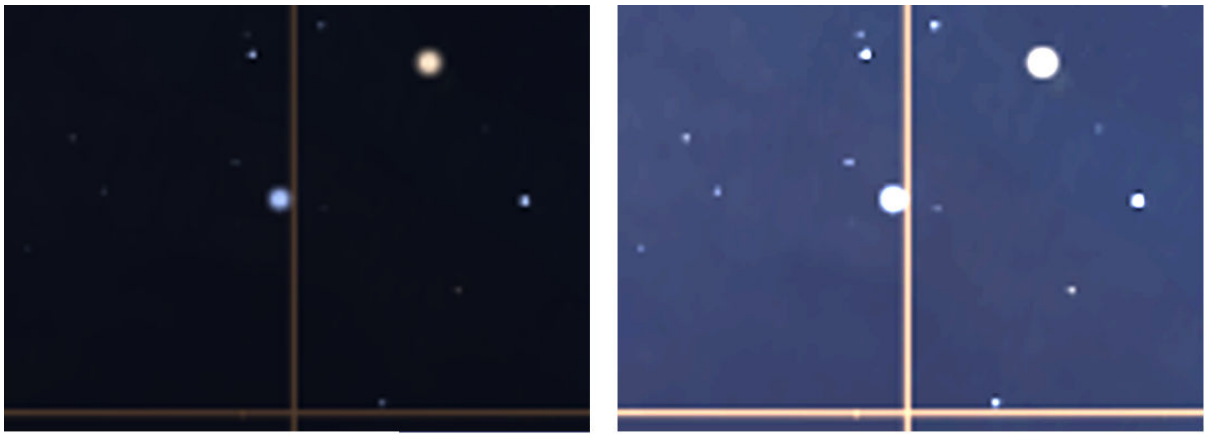

Figure 6: Around 2750 BC, in the simulation, Acrux is a dot near the line of the horizon.